\documentclass[conference,a4paper,romanappendices]{IEEEtran-v17}
\let\labelindent\relax % labelindent prevent from using enumitem, but is only there for legacy reasons, and hence we disable it.
\usepackage{enumitem}

\usepackage{ifthen}
\newboolean{arxiv}
\setboolean{arxiv}{true}

\ifCLASSINFOpdf
  \usepackage[pdftex]{graphicx}
\else
\fi
\usepackage[cmex10]{amsmath}

\usepackage[utf8]{inputenc}

\usepackage{amssymb}
\usepackage{bbm}
\usepackage{xspace}

\makeatletter
\def\@IEEEinterspaceratioM{0.265}
\def\@IEEEinterspaceMINratioM{0.1651}
\def\@IEEEinterspaceMAXratioM{0.38}

\def\@IEEEinterspaceratioB{0.31}
\def\@IEEEinterspaceMINratioB{0.19}
\def\@IEEEinterspaceMAXratioB{0.38}
\@IEEEtunefonts
\makeatother
\newtheorem{definition}{Definition}
\newtheorem{property}{Property}

\newtheorem{remark}{Remark}
\newtheorem{theorem}{Theorem}
\newtheorem{lemma}{Lemma}

\providecommand{\ee}[1]{\exp\mathopen{}\left(#1\right)}

\providecommand{\vectornorm}[1]{\left|\left|#1\right|\right|}

\providecommand{\indi}[1]{\mathbbm{1}\left\{#1\right\}}

\providecommand{\X}{\mathbf{X}}

\providecommand{\x}{\mathbf{x}}

\newcommand{\const}{\mathbbm{c}}

\newcommand{\E}[1]{\mathbb{E}\mathopen{}\left[#1\right]}
\newcommand{\inff}[1]{\inf\mathopen{}\left\{#1\right\}}

\newcommand{\supp}[1]{\sup\mathopen{}\left\{#1\right\}}
\newcommand{\minn}[1]{\min\mathopen{}\left\{#1\right\}}
\newcommand{\maxx}[1]{\max\mathopen{}\left\{#1\right\}}
\newcommand{\EE}[2]{\mathbb{E}_{#1}\mathopen{}\left[#2\right]}

\newcommand{\Va}[1]{\text{Var}\mathopen{}\left[#1\right]}
\newcommand{\Vaa}[2]{\text{Var}_{#1}\mathopen{}\left[#2\right]}

\newcommand{\farg}[1]{\mathopen{}\left( #1 \right)}
\newcommand{\bigoh}[1]{\mathcal{O}\mathopen{}\left(#1\right)}

\newcommand{\pr}[1]{\text{Pr}\mathopen{}\left[#1\right]}

\newcommand{\spacing}{\vspace*{0.5mm}}
\newcommand{\Msf}{M^*_{\text{sf}}}

\newcommand{\lefto}{\mathopen{}\left}

\newcommand{\sprob}{q}

\begin{document}
\title{Broadcasting a Common Message with Variable-Length Stop-Feedback Codes}

\author{\IEEEauthorblockN{Kasper Fløe Trillingsgaard\IEEEauthorrefmark{1},
Wei Yang\IEEEauthorrefmark{2},
Giuseppe Durisi\IEEEauthorrefmark{2}, and
Petar Popovski\IEEEauthorrefmark{1}}
\IEEEauthorblockA{\IEEEauthorrefmark{1} Aalborg University, 9220 Aalborg, Denmark}
\IEEEauthorblockA{\IEEEauthorrefmark{2} Chalmers University of Technology,
41296 Gothenburg, Sweden}}


%Telephone: (800) 555--1212, Fax: (888) 555--1212}


\maketitle

\begin{abstract}
We investigate the maximum coding rate achievable over a two-user broadcast channel for the scenario where a common message is transmitted using variable-length stop-feedback codes.
Specifically, upon decoding the common message, each decoder sends a stop signal to the encoder, which transmits continuously until it receives both stop signals.
For the point-to-point case, Polyanskiy, Poor, and Verdú (2011) recently demonstrated that variable-length coding combined with stop feedback significantly increases the speed at which the maximum coding rate converges to capacity.
This speed-up manifests itself in the absence of a square-root penalty in the asymptotic expansion of the maximum coding rate for large blocklengths, a result a.k.a. \emph{zero dispersion}.
In this paper, we show that this speed-up does not necessarily occur for the broadcast channel with common message. Specifically, there exist scenarios for which variable-length stop-feedback codes yield a positive dispersion.
\end{abstract}

\IEEEpeerreviewmaketitle

\section{Introduction}
We consider the setup where an encoder wishes to convey a common message over a broadcast channel with noiseless feedback to two decoders. Similarly to the single-decoder (SD) case, noiseless feedback combined with fixed-blocklength codes does not improve capacity, which is given by \cite[p. 126]{Gamal2011}
\begin{align}
  C = \sup_{P} \minn{I(P, W_1), I(P,W_2)}.\label{eq:common_capacity}
\end{align}
Here, $W_1$ and $W_2$ denote the channels to decoder $1$ and $2$, respectively, and the supremum is over all input distributions $P$.
For the case when there is no feedback, the speed at which $C$ is approached as the blocklength $n$ increases is of the order ${1}/{\sqrt{n}}$ \cite{Polyanskiy} (same as in the SD case). The constant factor associated to the ${1}/{\sqrt{n}}$ term is commonly referred to as channel \emph{dispersion}.

For the SD case, noiseless feedback combined with variable-length codes improve significantly the speed of convergence to capacity.
Specifically, it was shown in~\cite{Polyanskiy2011} that
\begin{align}
  \frac{1}{l}\log \widetilde M_{\text{f}}^*(l, \epsilon) = \frac{\widetilde C}{1-\epsilon} -\mathcal{O}\farg{\frac{\log l}{l}}\label{eq:SD_vlf_rate}
\end{align}
where $l$ stands for the average blocklength (average transmission time), $\widetilde M_{\text{f}}^*(l,\epsilon)$ is the maximum number of codewords in the SD case, and $\widetilde C$ denotes the corresponding capacity.
One sees from~\eqref{eq:SD_vlf_rate} that no square-root penalty occurs (zero dispersion), which implies a fast convergence to the asymptotic limit. This fast convergence is demonstrated numerically in \cite{Polyanskiy2011} by means of nonasymptotic bounds.
Variable-length stop-feedback (VLSF) codes, i.e., coding schemes where the feedback is used only to stop transmissions, are sufficient to achieve \eqref{eq:SD_vlf_rate}.

The purpose of this paper is to investigate whether a similar result holds for the broadcast channel with common message.

\subsubsection*{Contribution}
We consider the subclass of discrete memoryless broadcast channels for which $I(P,W_1)$ and $I(P,W_2)$ are maximized by the same input distribution $P^*$, which we assume to be unique. In this case, $C = \minn{I(P^*,W_1),I(P^*,W_2)}$. Focusing on the case when VLSF codes are used, we obtain nonasymptotic achievability and converse bounds on the maximum number of codewords $\Msf(l,\epsilon)$ with average blocklength $l$ that can be transmitted with reliability $1-\epsilon$. Here, the subscript ``sf'' stands for stop feedback.
By analyzing these bounds in the large-$l$ regime, we prove that when the two subchannels are independent and have the same capacity and the same dispersion, and when $\epsilon\leq 0.1968$, the asymptotic expansion of $\Msf(l,\epsilon)$ contains a square-root penalty (see \eqref{eq:asymp_expansion} and \eqref{eq:asymp_expansion_sym} for a precise statement of this result).
Hence, the fast convergence to the asymptotic limit experienced in the SD case cannot be expected.

The intuition behind this result is as follows: in the SD case, the stochastic variations of the information density that result in the square-root penalty can be virtually eliminated by using variable-length coding with stop-feedback.
Indeed, decoding is stopped after the information density exceeds a certain threshold, which yields only negligible stochastic variations. In the broadcast setup, however, the stochastic variations in the difference between the stopping times at the two decoders make the square-root penalty reappear.
Note that our result does not necessarily imply that feedback is useless. It only shows that VLSF codes cannot be used to speed-up convergence to the same level as in the SD case.
\subsubsection*{Proof techniques}
The achievability bound is an extension of~\cite[Th~3]{Polyanskiy2011}; the converse bound is based on an optimal stopping problem, where the probability that the stopping time exceeds a given threshold is minimized under a constraint on the ``stopped'' information density process.
The asymptotic analysis of the converse bound relies on Hoeffding's inequality and on the Berry-Esseen central limit theorem, whereas the asymptotic analysis  of the achievability bound relies on asymptotic results for random walks~\cite{Gut2009} and on a Berry-Esseen-type theorem that holds for random summations~\cite{Landers1976}.

\subsubsection*{Notation}
Upper case, lower case, and calligraphic letters denote random variables (RV), deterministic quantities, and sets, respectively.
The probability density function of a standard Gaussian RV  is denoted by  $\phi(x)$.
Furthermore, $\Phi(x)\triangleq 1- Q(x)$ is its cumulative distribution, where $Q(x)$ is the Q-function.
We let $x^+$ and $x^-$ denote $\max(0,x)$ and $\minn{0,x}$, respectively.
Throughout the paper, the index $k$ belongs always to the set $\{1,2\}$, although this is sometimes omitted.
Furthermore, $\bar k \triangleq 3- k$. We adopt the convention that $\sum_{i=j}^{j-1} a_i = 0$ for all $\{a_i\}$ and all integers $j$.
We use ``$\const$'' to denote a finite nonnegative constant. Its value may change at each occurrence.
 Finally, $\mathbb{N}$ denotes the set of positive integers and $\mathbb{Z}_+ = \mathbb{N}\cup \{0\}$.

\section{System Model}
A common-message discrete memoryless broadcast channel with two decoders is defined by the finite input alphabet $\mathcal{X}$ and the finite output alphabets $\mathcal{Y}_k$, along with the stochastic matrices $W_k$, where $W_k(y_k|x)$ denotes the probability that $y_k\in\mathcal{Y}_k$ is observed at decoder $k$ given $x\in\mathcal{X}$. We assume that the outputs at each time~$i$ are conditionally independent given the input, i.e.,
\begin{align}\label{eq:factorization}
  P_{Y_{1,i},Y_{2,i}|X_i}(y_{1,i},y_{2,i}|x_i) \triangleq W_1(y_{1,i}|x_i)W_2(y_{2,i}|x_i).
\end{align}
Define the set of probability distributions on $\mathcal{X}$ by $\mathcal{P}(\mathcal{X})$.
Let $P \times W_k: (x,y_k) \rightarrow P(x)W(y_k|x)$ denote the joint distribution of input and output at decoder $k$, and let $PW_k: y_k \rightarrow \sum_{x\in\mathcal{X}} P(x)W_k(y_k | x)$ denote the marginal distribution on $\mathcal{Y}_k$.
For every $P\in\mathcal{P}(\mathcal{X})$, the information density is defined as
\begin{align}
  \imath_{P, W_k}(x^n; y_k^n) \triangleq\sum_{i=1}^n \log \frac{W_k(y_{k,i}|x_i)}{PW_k(y_{k,i})}.
\end{align}
We let $I(P, W_k)\triangleq \EE{P\times W_k}{\imath_{P, W_k}(X; Y_k)}$ be the mutual information, $V(P, W_k)\triangleq \Vaa{P\times W_k}{\imath_{P, W_k}(X; Y_k)}$ be the (unconditional) information variance, and $T(P,W_k)\triangleq \EE{P\times W_k}{|\imath_{P, W_k}(X; Y_k) - I(P,W_k)|^3}$ be the third absolute moment of the information density.
We restrict ourselves to the case, where there exists a unique probability distribution $P^*\in\mathcal{P}(\mathcal{X})$ that maximizes simultaneously both $I(P,W_1)$ and $I(P,W_2)$.
In this case, the capacity is given by
\begin{align}
C \triangleq \min\{ C_1,C_2\}\label{eq:capacity_def}
\end{align}
where $C_k\triangleq I(P^*, W_k)$.
The corresponding (unique) capacity-achieving output distributions are denoted by $P_{Y_k}^*$. %, are
% \begin{align}
%   P_{Y_k}^*(y_k) &\triangleq P^*W_k(y_k).
% \end{align}
Finally, we also define the dispersions $V_{k} \triangleq V(P^*,W_k)$.
% \begin{align}
% V_{k} \triangleq V(P^*,W_k).
% \end{align}

We are now ready to formally define a VLSF code for the broadcast channel with common message.
\begin{definition}
\label{def:VLSFcode}
An $(l, M,\epsilon)$-VLSF code for the broadcast channel with common message consists of:
\begin{enumerate}[leftmargin=*]
\item A RV $U\in\mathcal{U}$, with $|\mathcal{U}|\leq 3$, which is known by the encoder and by both decoders.
\item A sequence of encoders   $f_n: \mathcal{U}\times \mathcal{M} \rightarrow \mathcal{X}$, each one mapping the message $J\in\mathcal{M}=\{1,\ldots,M\}$, drawn uniformly at random, to the channel input according to $X_n = f_n(U,J)$.
\item Two nonnegative integer-valued RVs $\tau_1$ and $\tau_2$ that are stopping times with respect to the filtrations $\mathcal{F}(U, Y_{1}^n)$ and $\mathcal{F}(U, Y_{2}^n)$, respectively, and which satisfy
\begin{align}
  \E{\maxx{\tau_1, \tau_2}}\leq l.\label{eq:avg_blocklength_const}
\end{align}
\item A sequence of decoders $g_{k,n}: \mathcal{U}\times \mathcal{Y}_i^n \rightarrow \mathcal{M}$ satisfying
\begin{align}
  \pr{J \not= g_{k,\tau_k}(U, Y_k^{\tau_k}) } \leq \epsilon, \qquad k\in\{1,2\}.\label{eq:def_prob_error}
\end{align}
\end{enumerate}
\end{definition}
\begin{remark}
The RV $U$ serves as common randomness, and enables the use of randomized codes~\cite{Lapidoth1998}.
To establish the cardinality bound on $U$, we proceed as in \cite[Th.~19]{Polyanskiy2011} to show that $|\mathcal{U}|\leq 4$ is sufficient. This bound can be further improved to $|\mathcal{U}|\leq 3$ by using the Fenchel-Eggleston theorem \cite[p.~35]{Eggleston}.
\end{remark}
\begin{remark}
VLSF codes require a feedback link from the decoders to the encoder.
This feedback consists of a $1$-bit stop signal per decoder, which is sent by decoder $k$ at time~$\tau_k$. The encoder continuously transmits until both decoders have fed back a stop signal.
Hence, the blocklength is  $\maxx{\tau_1,\tau_2}$.
%Note that the stopping times can not be computed by the encoder without full noiseless feedback.
\end{remark}
%\spacing

Our aim is to characterize the largest number of codewords $\Msf(l, \epsilon)$, whose average length is $l$, that can be transmitted with reliability $1-\epsilon$ using a VLSF code.
% Mathematically,
% \begin{align}
%   \Msf(l, \epsilon) &= \maxx{M : \exists (l,M,\epsilon)\text{-VLSF code}}.
% \end{align}

\section{Main Results}
% We present an achievability bound based on \cite[Th. 3]{Polyanskiy2011}, and a converse bound based on the meta-converse theorem~\cite[Th.~26]{Polyanskiy2010b}.
% Based on these bounds, we obtain an asymptotic expansion of $\Msf(l,\epsilon)$.
\subsection{Achievability bound}
We first present an achievability bound.
Its proof (omitted) follows  closely the proof of~\cite[Th. 3]{Polyanskiy2011}.
\begin{theorem}
\label{thm:simple_achiev}
  Fix $P \in\mathcal{P}(\mathcal{X})$. Let $\gamma_1,\gamma_2\geq 0$ and $0 \leq \sprob \leq 1$  be arbitrary scalars. Let the stopping times $\tau_k$ and $\bar \tau_k$, $k\in\{1,2\}$, be defined as
  \begin{align}
  \tau_k &\triangleq \inff{n \geq 0: \imath_{P, W_k}(X^n ; Y_k^n)\geq \gamma_k}\\
  \bar \tau_k &\triangleq \inff{n \geq 0: \imath_{P, W_k}(\bar X^n ; Y_k^n)\geq \gamma_k}
  \end{align}
  where $(X^n, \bar X^n, Y_1^n, Y_2^n)$ are jointly distributed according to
  \begin{align}
    &P_{X^n, \bar X^n,Y_1^n, Y^n_2}(x^n, \bar x^n, y_1^n, y_2^n) \nonumber\\
    &= P_{Y_1^n, Y^n_2|X^n}(y_1^n, y_2^n|x^n)\prod_{i=1}^n P(x_i)P(\bar x_i).
  \end{align}
For every $M$, there exists an $(l, M, \epsilon)$-VLSF code such that
\begin{align}
    l &\leq (1-\sprob) \E{\maxx{\tau_1,\tau_2}}\label{eq:achiev_Emax}
\end{align}
    and
\begin{align}
    \epsilon &\leq  \sprob + (1-\sprob)(M-1)\pr{\tau_k\geq \bar\tau_k}.\label{eq:achiev_Proberror1}
\end{align}
\end{theorem}
\begin{remark}
\label{rem:prob_error}
Following the same steps as in~\cite[Eq. (111)--(118)]{Polyanskiy2011},
  $\epsilon$ in~\eqref{eq:achiev_Proberror1} can be further upper-bounded as
  \begin{align}
  \epsilon \leq \sprob + (1-\sprob)(M-1)\exp\left\{-\gamma_k\right\}.\label{eq:achiev_Proberror2}
  \end{align}
   This  bound is easier to evaluate and to analyze asymptotically.
\end{remark}

\subsection{Converse bound}
Let $P_{\x^n}\in\mathcal{P}(\mathcal{X})$ be the type~\cite[Def. 2.1]{Csiszar} of the sequence $\x^n \in\mathcal{X}^n$. 
We are now ready to state our converse bound.
\begin{theorem}
\label{thm:converse_bound}
For every $M$, $t\in \mathbb{Z}_+$ and $\delta>0$, let
\begin{align}
\lambda_t\triangleq \log M - \log \log M -\delta - (|\mathcal{X}|-1)\log(t+1)\label{eq:conve_lambda_def}
\end{align}
and let
\begin{multline}
  L_t \triangleq \prod_{k=1}^2 \max_{ \x^t\in\mathcal{X}^t}\left\{\pr{\imath_{P_{\x^t},W_k}(\x^t;Y_k^t)> \lambda_t}\right\} \\
  + \varepsilon_M \left(1+ \min_{k} \max_{ \x^t\in\mathcal{X}^t} \pr{\imath_{P_{\x^t},W_k}(\x^t;Y_k^t)> \lambda_t} \right)\label{eq:Lt_def}
\end{multline}
where $\varepsilon_M = \epsilon + (\log M)^{-1}$.
Then, for every $(l,M,\epsilon)$-VLSF code, we have
\begin{align}
  l \geq \sum_{t=0}^{\infty}\left(1- L_t\right)^+.\label{eq:conve_l_lower_bound}
\end{align}
\end{theorem}
\begin{IEEEproof}
  See Section~\ref{sec:converse_bound}.
\end{IEEEproof}

\subsection{Asymptotic expansion}
Analyzing~\eqref{eq:achiev_Proberror2} and~\eqref{eq:conve_l_lower_bound} in the  limit $l\to\infty$, we obtain the following asymptotic characterization of $\Msf(l,\epsilon)$.
\begin{theorem}
\label{thm:asymp}
Let $Z_k \sim \mathcal{N}(0,1)$, $V = \sqrt{V_1 V_2}$, $\varrho_k = \left({V_k}/{V_{\bar k}}\right)^{1/4}$, and let $y=\tilde Q^{-1}\farg{x}$ be the solution of
\begin{align}
 \prod_{k=1}^2 Q\farg{ -\varrho_k y}+ x \left(1 + \min_{k} Q\farg{-\varrho_k y}\right) = 1.\label{eq:tildeQ_def}
\end{align}
For every discrete memoryless broadcast channel with $C_1=C_2$ and every $\epsilon \in (0,1)$, we have
\begin{multline}
   \frac{C l}{1-\epsilon}-  \Xi_{\text{a}} \sqrt{l}-\mathcal{O}\farg{ l^{1/4+\delta} } \leq \log \Msf(l,\epsilon)  \\
  \leq \frac{C l}{1-\epsilon}- \Xi_{\text{c}}\sqrt{l } +\mathcal{O}(\log l) \label{eq:asymp_expansion}
\end{multline}
where $\delta>0$ is an arbitrarily small constant,
\begin{align}\label{eq:constant_LHS_as}
  \Xi_{\text{a}} \triangleq \sqrt{\frac{V_1 + V_2}{2\pi(1-\epsilon)}}
\end{align}
and
\begin{multline}\label{eq:constant_RHS_as}
  \Xi_{\text{c}} \triangleq \sqrt{\frac{V}{(1-\epsilon)^3}}\left(\E{\minn{\tilde Q^{-1}\farg{\epsilon},\max_{k}\varrho_k Z_k} }\right.\\
  \left.-\epsilon \left(2\tilde Q^{-1}(\epsilon) -\min_{k} \E{\minn{\tilde Q^{-1}\farg{\epsilon}, \varrho_k Z_k} }   \right)\right).
\end{multline}
\end{theorem}
\begin{IEEEproof}
  The converse bound in \eqref{eq:asymp_expansion} is proved in Section~\ref{sec:converse_asymptotics} and the achievability bound is proved in Section~\ref{sec:achievability_asymptotics}.
\end{IEEEproof}

\begin{remark}
When $C_1\not= C_2$, it can be shown that the square-root penalty on the LHS of \eqref{eq:asymp_expansion} vanishes. In this case, the problem reduces to the point-to-point transmission to the weakest decoder, for which the zero-dispersion result in \cite{Polyanskiy2011} applies.
\end{remark}
\begin{remark}
  For the case when $P_{Y_{1,i}, Y_{2,i} | X_i}$ does not satisfy~\eqref{eq:factorization}, a bound similar to the LHS of~\eqref{eq:asymp_expansion} can be obtained by replacing $\Xi_\text{a}$ in~\eqref{eq:constant_LHS_as} with
  \begin{align}
   \sqrt{\frac{V_1 + V_2 - 2\text{Cov}(\imath_{P^*,W_1}(X; Y_1),\imath_{P^*,W_2}(X; Y_2))}{2\pi(1-\epsilon)}}.
  \end{align}
\end{remark}
\begin{remark}
When $\varrho_1=\varrho_2=1$ (and, hence, $V_1=V_2$), one can simplify the RHS of~\eqref{eq:asymp_expansion} as follows:
\begin{align}
&\log \Msf(l,\epsilon)\nonumber\leq \frac{C l }{1-\epsilon} - \sqrt{\frac{V l}{(1-\epsilon)^3}}\nonumber\\
 &\quad\times\bigg( \frac{1}{\sqrt{\pi}}\left(1- Q\farg{\sqrt{2}Q^{-1}(\epsilon)} \right)+ (\epsilon - 2)\phi\farg{Q^{-1}(\epsilon)} \bigg) \nonumber\\
  & \quad- \mathcal{O}\farg{\log l}.\label{eq:asymp_expansion_sym}
\end{align}
The second-order term in~\eqref{eq:asymp_expansion_sym} is strictly negative for all $\epsilon \leq 0.1968$. This implies that, when $C_1=C_2$, $V_1=V_2$, and $\epsilon \leq 0.1968$, the asymptotic expansion of $\log \Msf(l,\epsilon)$ contains a square-root penalty.

\end{remark}

\section{Proof of Theorem~\ref{thm:converse_bound}}
\label{sec:converse_bound}
Fix $M$ and $\epsilon$. To establish Theorem~\ref{thm:converse_bound}, we derive a lower bound on $l$ that holds for all VLSF codes having $M$ codewords and probability of error no larger than $\epsilon$. Since,
\begin{align}
  l\geq \E{\maxx{\tau_1, \tau_2}} = \sum_{t=0}^\infty \left(1 - \pr{\maxx{\tau_1, \tau_2}\leq t}\right)\label{eq:conve_Etau1_tau2_in_terms_of_pr_tau1_tau_2}
\end{align}
we can lower-bound $l$ by upper-bounding $\pr{\maxx{\tau_1,\tau_2}\leq t}$ for every $t\in\mathbb{Z}_+$. The following property \ifthenelse{\boolean{arxiv}}{(proven in Appendix~\ref{app:conve_property})}{} turns out to be useful.
\begin{property}
 Fix $t\in\mathbb{Z}_+$ and $\alpha\in[0,1]$, and suppose there exists an $(l,M,\epsilon)$-VLSF code with $\pr{\maxx{\tau_1,\tau_2} \leq t}\leq \alpha$. Then there exists an $(l',M,\epsilon)$-VLSF code for some $l' \geq l$, for which $\pr{\maxx{\tau_1,\tau_2} \leq t}\leq \alpha$ and $\tau_1,\tau_2\in\{t,t+1,\ldots\}$.
  \label{prop:VLSF_t}
\end{property}
\spacing

Fix  an arbitrary $(l, M,\epsilon)$-VLSF code, defined by the tuple $(f_n,g_{1,n},g_{2,n},\tau_1,\tau_2,U)$. By Property~\ref{prop:VLSF_t}, it is sufficient to consider codes for which $\tau_1,\tau_2\in\{t,t+1,\cdots\}$. Let $\epsilon_{k}^{(u)}$, $u\in\mathcal{U}$, be constants in $[0,1]$ such that $\sum_{u\in\mathcal{U}}P_U(u) \epsilon_k^{(u)}\leq \epsilon$ and $\pr{J\not= g_{k,\tau_k}(U, Y_k^{\tau_k})|U=u}\leq \epsilon_k^{(u)}$.
%Such constants exist by \eqref{eq:def_prob_error}.

Since  $\{\tau_k = n\} \in \mathcal{F}\farg{U, Y_k^n}$, we can define a sequence of binary functions $\varphi_k \triangleq \{\varphi_{k,t},\varphi_{k,t+1},\cdots\}$ such that $\varphi_{k,n}(u,y^n_k) \triangleq\indi{\tau_k = n}$.
Let $P^{(u)}_{\X}$ be the conditional probability measure on $\mathcal{X}^\infty$ induced by the encoder given $U=u$.
Define for $u\in\mathcal{U}$ the set $\mathcal{\bar Y}^{(u)}_k \triangleq \left\{ y^n\in\mathcal{Y}^n_k : \varphi_{k,n}(u,y^n) = 1 \right\}$.
Note that we must have $Y_k^{\tau_k}\in \mathcal{\bar Y}^{(u)}$.
Let the length of a sequence of channel outputs $\bar y\in \mathcal{\bar Y}_k^{(u)}$ be denoted by $|\bar y|$.
On $\mathcal{\bar Y}_k^{(u)}$, define the conditional probability measure $\mathbbm{P}^{(k,u)}_{\bar Y | \X}$, given $\x \in \mathcal{X}^\infty$ and $u\in\mathcal{U}$, as
\begin{align}
\mathbbm{P}^{(k,u)}_{\bar Y |\X}(\bar y | \x) \triangleq \prod_{i=1}^{|\bar y|}W(\bar y_i |\x_i)
\end{align}
and the probability measure $\mathbbm{P}^{(k,u)}_{\bar Y,\X}(\bar y, \x)\triangleq \mathbbm{P}^{(k,u)}_{\bar Y |\X}(\bar y | \x) P^{(u)}_{\X}(\x)$ on $\mathcal{\bar Y}^{(u)}\times \mathcal{X}^\infty$. We also need the following auxiliary probability measure $\mathbbm{Q}^{(k,u)}_{\bar Y}$ on $\mathcal{\bar Y}_k^{(u)}$
\begin{align}
  &\mathbbm{Q}^{(k,u)}_{\bar Y}\farg{\bar y} \triangleq\nonumber\\
    &\sum_{P_{\x^{t}} \in \mathcal{P}_{t}(\mathcal{X})} \bigg(\frac{1}{|\mathcal{P}_{t}(\mathcal{X})|} \prod_{i=1}^{t} P_{\x^t}W_k(\bar y_{i}) \prod_{i=t+1}^{|\bar y|} P_{Y_k}^*\farg{\bar y_{i}}\bigg)\label{eq:Qk_def}
\end{align}
and the probability measure $\mathbbm{Q}^{(k,u)}_{\bar Y, \X}\farg{\bar y,\x}=\mathbbm{Q}^{(k)}_{\bar Y}\farg{\bar y} P^{(u)}_{\X}(\x)$ on $\mathcal{\bar Y}^{(u)}\times \mathcal{X}^\infty$. Here, $\mathcal{P}_t(\mathcal{X})\subseteq \mathcal{P}(\mathcal{X})$ denotes the set of types formed by length-$t$ sequences.

Using the meta-converse theorem \cite[Th.~27]{Polyanskiy2010b}, the inequality \cite[Eq. (102)]{Polyanskiy2010b}, the fact that $\mathbbm{Q}^{(k,u)}_{\bar Y_k, \X}$ is a convex combination of distributions~\cite[Lem. 3]{Tomamichel2013}, and the upper bound $|\mathcal{P}_t(\mathcal{X})| \leq (t+1)^{|\mathcal{X}|-1}$ \cite[Lem. 1.1]{Tan2014}, we conclude that \ifthenelse{\boolean{arxiv}}{(see Appendix~\ref{app:conve_meta_converse})}{(see details in \cite[App.~I-B]{Trillingsgaard2015arxiv})}
\begin{align}
 \mathbbm{P}_{\bar Y,\X}^{(k,u)}\mathopen{}\left[\tilde \imath_{k}^{(u)}(\X ; \bar Y_k) \leq \lambda_t\right] \leq \varepsilon_{k,M}^{(u)}\label{eq:conve_final_const}
\end{align}
where $\varepsilon_{k,M}^{(u)} \triangleq \epsilon_k^{(u)} + (\log M)^{-1}$ and $\lambda_t$ is defined in \eqref{eq:conve_lambda_def}. Here,
\begin{align}
  \tilde \imath_{k}^{(u)}(\x;\bar y) \triangleq \imath_k(\x^t ; y^t)+\sum_{i=t+1}^{|\bar y|} \log \frac{W_k(y_{i} | \mathbf{x}_i)}{P_{Y_k}^*(y_{i})}
\end{align}
where $\imath_k(\x^t ; y^t) \triangleq \imath_{P_{\x^t},W_k}(\x^t, y^t)$. Next, we minimize $\pr{\tau_k \leq t | U =u}$ over all stopping times $\tau_k$ satisfying \eqref{eq:conve_final_const}:
 \begin{IEEEeqnarray}{rCL}
   \IEEEeqnarraymulticol{3}{l}{\pr{\tau_k \leq t | U =u}=\mathbbm{P}_{\bar Y,\X}^{(k,u)}\mathopen{}\left[|\bar Y| = t\right]}\nonumber\\
  &=&  \mathbbm{P}_{\bar Y,\X}^{(k,u)}\mathopen{}\left[\tilde \imath_{k}^{(u)}(\X;\bar Y_k)> \lambda_t,|\bar Y| = t\right] \nonumber\\
  &&+ \mathbbm{P}_{\bar Y,\X}^{(k,u)}\mathopen{}\left[\tilde \imath_{k}^{(u)}(\X;\bar Y_k)\leq \lambda_t,|\bar Y| = t\right]\\
  &\leq& \min\Big\{1, \mathbbm{P}_{\bar Y,\X}^{(k,u)}\lefto[\tilde \imath_{k}^{(u)}(\X;\bar Y_k)> \lambda_t,|\bar Y| = t\right] + \varepsilon_{k,M}^{(u)}\Big\}\IEEEeqnarraynumspace\label{eq:tau_k_opt}\\
  &\leq& \max_{ \x^t\in\mathcal{X}^t}\pr{\imath_k(\x^t;Y_k^t)> \lambda_t} \nonumber\\
  &&+ \minn{\varepsilon_{k,M}^{(u)},1-\max_{ \x^t\in\mathcal{X}^t}\pr{\imath_k(\x^t;Y_k^t)> \lambda_t}}.\label{eq:conve_sep_decoders_res}
 \end{IEEEeqnarray}
 Here, \eqref{eq:tau_k_opt} follows from \eqref{eq:conve_final_const}.
 %
%
% \begin{align}
%   &\mathbbm{P}_{\bar Y,\X}^{(k,u)}\mathopen{}\left[|\bar Y_k| = t\right]\nonumber\\
%   &= \mathbbm{P}_{\bar Y,\X}^{(k,u)}\mathopen{}\left[\tilde \imath_{k}^{(u)}(\X;\bar Y_k)> \lambda_t,|\bar Y_k| = t\right] \nonumber\\
%   &\qquad\qquad+ \mathbbm{P}_{\bar Y,\X}^{(k,u)}\mathopen{}\left[\tilde \imath_{k}^{(u)}(\X;\bar Y_k)\leq \lambda_t,|\bar Y_k| = t\right]\\
%    &\leq \min\Big\{1, \mathbbm{P}_{\bar Y,\X}^{(k,u)}\mathopen{}\left[\tilde \imath_{k}^{(u)}(\X;\bar Y_k)> \lambda_t,|\bar Y_k| = t\right] + \varepsilon_{k,M}^{(u)}\Big\}\\
%   &\leq \max_{ \x^t\in\mathcal{X}^t}\pr{\imath_k(\x^t;Y_k^t)> \lambda_t} \nonumber\\
%   &\quad+ \minn{\varepsilon_{k,M}^{(u)},1-\max_{ \x^t\in\mathcal{X}^t}\pr{\imath_k(\x^t;Y_k^t)> \lambda_t}}.\label{eq:conve_sep_decoders_res}
% \end{align}
Since the stopping times $\tau_1$ and $\tau_2$ are conditional independent given $U=u$,  \eqref{eq:conve_sep_decoders_res} implies that
\begin{align}
&\pr{\maxx{\tau_1, \tau_2} \leq t| U=u}= \prod_{k=1}^2 \mathbbm{P}_{\bar Y , \X}^{(k,u)}\mathopen{}\left[|\bar Y_k|=t\right]\\
    &\leq \prod_{k=1}^2 \max_{ \x^t\in\mathcal{X}^t}\left\{\pr{\imath_k(\x^t;Y_k^t)> \lambda_t}\right\} \nonumber\\
  &\quad+ \min_{k}\left\{ \varepsilon_{\bar k,M}^{(u)} +\varepsilon_{k,M}^{(u)}\max_{ \x^t\in\mathcal{X}^t}\pr{\imath_k(\x^t;Y_{\bar k}^t)> \lambda_t}\right\}.\label{eq:conve_maxtau1_tau2_u}
\end{align}
Note that \eqref{eq:conve_maxtau1_tau2_u} holds for all $\tau_k$ that satisfies \eqref{eq:conve_final_const}.
Averaging \eqref{eq:conve_maxtau1_tau2_u} over $u\in\mathcal{U}$ and using the inequality $\sum_{u\in\mathcal{U}}P_U(u) \varepsilon_{k,M}^{(u)}\leq \epsilon + {(\log M)^{-1}} = \varepsilon_M$, we obtain~\eqref{eq:Lt_def}. The proof is concluded using \eqref{eq:conve_Etau1_tau2_in_terms_of_pr_tau1_tau_2}.

\section{Asymptotic Analysis: Converse Bound}
\label{sec:converse_asymptotics}
We analyze $L_t$ in~\eqref{eq:Lt_def} in the limit $l\to\infty$.
By~\eqref{eq:conve_l_lower_bound},
\begin{align}
l \geq \sum_{t=0}^{\infty} (1- L_t)^+  \geq \sum_{t=0}^{\lfloor \beta \rfloor}(1- L_t)^+ \geq \sum_{t=0}^{\lfloor \beta \rfloor}(1- L_t)\label{eq:conve_proof_l_lower_bound_beta}
\end{align}
where $\beta>0$ will be specified shortly.
Let $\lambda\triangleq  \log M - \log \log M - \delta -(|\mathcal{X}|-1)\log(\beta + 1)$. For all $t\leq \beta$,
\begin{IEEEeqnarray}{rCl}
\max_{\x^t \in\mathcal{X}^t}\pr{\imath_k(\x^t;Y_k^t)> \lambda_t} &\leq& \max_{\x^t \in\mathcal{X}^t}\pr{\imath_k(\x^t;Y_k^t)> \lambda}.\IEEEeqnarraynumspace
\end{IEEEeqnarray}
The key step is to establish an asymptotic upper bound on $\max_{\x^t \in\mathcal{X}^t}\pr{\imath_k(\x^t;Y_k^t)> \lambda}$ for every $t\in\mathbb{Z}_+$ as $\lambda \rightarrow \infty$.
%With such upper-bound, \eqref{eq:conve_proof_l_lower_bound_beta} provides a lower bound on $l$.

Let $\alpha\triangleq\frac{\lambda}{C}-\sqrt{\frac{V\lambda}{C^3}} \log \lambda$ and let $\beta$ be the solution of
\begin{align}
  {(\lambda - \beta C)}/{\sqrt{\beta V }} = -\tilde Q^{-1}(\epsilon)\label{eq:beta_def}
\end{align}
where $C$ is given in \eqref{eq:capacity_def}, $V$ is defined in Theorem~\ref{thm:asymp}, and  $\tilde Q^{-1}(\epsilon)$ in~\eqref{eq:tildeQ_def}.
We divide the asymptotic analysis of $\max_{\x^t \in\mathcal{X}^t}\pr{\imath_k(\x^t;Y_k^t)> \lambda}$ into three cases: the ``large deviations regime" $t\in[0, \alpha)$, where we use Hoeffding's inequality, the ``central regime" $t\in[\alpha, \beta)$, where  Berry-Esseen central limit theorem is applied, and the case $t\geq \beta$, where the trivial upper bound  $\max_{\x^t \in\mathcal{X}^t}\pr{\imath_k(\x^t;Y_k^t)> \lambda}\leq 1$ suffices.

In the first case, invoking Hoeffding's inequality \cite[Th.~2]{Hoeffding1963} and using that $I(P_{\x^t},W_k)$ is upper-bounded by $C$ uniformly, we obtain \ifthenelse{\boolean{arxiv}}{(see Appendix~\ref{app:asymp_conve_hoeffding} for details)}{(see \cite[App.~II-A]{Trillingsgaard2015arxiv})}
\begin{align}
  \sum_{t=0}^{\lfloor \alpha \rfloor} \max_{\x^t \in\mathcal{X}^\infty}\pr{\imath_k(\x^t ; Y_k^t) > \lambda} = o(1), \quad \lambda\rightarrow \infty \label{eq:conve_sum_hoeffding_bound}
\end{align}
and
\begin{IEEEeqnarray}{rCl}
\sum_{t=0}^{\lfloor \alpha \rfloor} \prod_{k=1}^2 \max_{\x^t \in\mathcal{X}^t}\left\{\pr{\imath_k(\x^t ; Y_k^t) > \lambda} \right\}= o(1), \,\, \lambda\rightarrow \infty.\IEEEeqnarraynumspace \label{eq:conve_sum_hoeffding_bound_2}
\end{IEEEeqnarray}
%as $\lambda\rightarrow \infty$.

In the central regime, we use the Berry-Esseen central limit theorem \cite[Th.~V.3]{V.V.Petro} to show that
\begin{align}
&\pr{\imath_k(\x^t ; Y_k^t) > \lambda}\leq Q\farg{\frac{\lambda-t I(P_{\x^t},W_k)}{\sqrt{t V(P_{\x^t},W_k)}}}+ \frac{\const}{\sqrt{t}}.\label{eq:conve_berry}
\end{align}
We next maximize~\eqref{eq:conve_berry} over $\x^t\in\mathcal{X}^t$ following the approach in \cite[Prop. 8]{Tomamichel2013}.
Specifically, we use continuity properties of $I(P, W_k)$ and $V(P , W_k)$ for probability distributions $P\in\mathcal{P}(\mathcal{X})$ close to $P^*$ to show that \ifthenelse{\boolean{arxiv}}{(see Appendix~\ref{app:asymp_conve_berry})}{(see \cite[App. II-B]{Trillingsgaard2015arxiv})}
\begin{align}
    &\sum_{t=\lfloor \alpha \rfloor+1}^{\lfloor \beta \rfloor}\max_{\x^t \in \mathcal{X}^t } \pr{\imath_k(\x^t ; Y_k^t) > \lambda}\nonumber\\
        &\leq \sqrt{\frac{V\lambda}{C^3}}\Big(\tilde Q^{-1}\farg{\epsilon} - \E{\minn{ \tilde Q^{-1}\farg{\epsilon} , \varrho_k Z_k }}\Big) + \mathcal{O}(\log \lambda)\label{eq:conve_sum_berry_esseen_Q}
\end{align}
where $\varrho_k$ are defined in Theorem~\ref{thm:asymp} and $Z_k\sim \mathcal{N}(0,1)$.
Similarly, we obtain
\begin{align}
&\sum_{t=\lfloor \alpha \rfloor+1}^{\lfloor \beta \rfloor} \prod_{k=1}^2 \max_{\x^t \in \mathcal{X}^t}\pr{\imath_k(\x^t ; Y_k^t) > \lambda}\nonumber\\
&\leq   \sqrt{\frac{V\lambda}{C^3}}\left(\tilde Q^{-1}\farg{\epsilon} - \E{\minn{ \tilde Q^{-1}\farg{\epsilon} , \max_k \varrho_k Z_k}} \right) \nonumber\\
& \quad+ \mathcal{O}(\log \lambda).\label{eq:conve_sum_berry_esseen_Q2}
\end{align}
Using \eqref{eq:conve_proof_l_lower_bound_beta}, \eqref{eq:conve_sum_hoeffding_bound}, \eqref{eq:conve_sum_hoeffding_bound_2}, \eqref{eq:conve_sum_berry_esseen_Q}, and \eqref{eq:conve_sum_berry_esseen_Q2}, we obtain
\begin{IEEEeqnarray}{rCl}
  l &\geq& \sum_{t=0}^{\lfloor \beta \rfloor} (1-L_t)\\
    &\geq&  \frac{\lambda(1-\varepsilon_M)}{C} +  \sqrt{\frac{V\lambda}{C^3}} \bigg( \E{ \minn{\tilde Q^{-1}\farg{\epsilon}, \max_{k} \varrho_k Z_k }}\nonumber\\
   &&-\varepsilon_M\bigg( 2\tilde Q^{-1}(\epsilon)-   \min_{k} \E{ \minn{\tilde Q^{-1}\farg{\epsilon} , \varrho_k Z_k }}\bigg) \bigg)
    \nonumber\\
    &&-\mathcal{O}(\log \lambda)\label{eq:conve_int_form}
\end{IEEEeqnarray}
as $\lambda \rightarrow \infty$.
Finally, we have that
\begin{IEEEeqnarray}{rCL}
 \lambda &=&\log M -\log\log M-\delta - (|\mathcal{X}| -1)\log(\beta+1) \\
 &\leq& \frac{C l }{1-\varepsilon_M} \nonumber\\
 &&- \sqrt{\frac{V l}{(1-\varepsilon_M)^3}}\left(\E{\minn{ \tilde Q^{-1}\farg{\epsilon},\max_{k}\varrho_k Z_k} }\right.\nonumber\\
  &&-\left.\varepsilon_M \left(2\tilde Q^{-1}(\epsilon) -\min_{k} \E{\minn{\tilde Q^{-1}\farg{\epsilon},\varrho_k Z_k} }   \right)\right) \nonumber\\
  &&+ \mathcal{O}\farg{\log l}\label{eq:conve_res_before_taylor}
\end{IEEEeqnarray}
as $l\rightarrow \infty$. The final result in~\eqref{eq:asymp_expansion} is obtained through algebraic manipulations.

\section{Asymptotic Analysis: Achievability Bound}
\label{sec:achievability_asymptotics}
% We intend to invoke Theorem~\ref{thm:simple_achiev}, and hence we need to fix an input distribution $P\in\mathcal{P}(\mathcal{X})$, a RV $\Coopvar$ and thresholds $\gamma_k^{(\coopvar)}$, $\coopvar\in\{1,2\}$.
Set $P=P^*$, and fix $r\in\mathbb{N}$, $\sprob = \frac{l' \epsilon -1}{ l' - 1}$, and $l'>0$, a parameter that will be related to the average blocklength. 
%Choose $\Coopvar\in\{1,2\}$ such that $\Coopvar=1$ with probability $\frac{l' \epsilon -1}{ l' - 1}$ and $\Coopvar=2$ with probability $\frac{l' (1-\epsilon)}{ l' - 1}$. 
Let the thresholds be chosen as follows: 
 \begin{align}
  \gamma \triangleq \gamma_k\triangleq C \left( l' - g(C l') \right).\label{eq:gamma_def}
\end{align}
Here,
\begin{align}
g\farg{x}\triangleq \sqrt{\frac{ V_1+V_2}{2\pi C^2}}\sqrt{ \frac{x}{C}}+b_1 x^{\frac{r+1}{4r+2}} \log x
\end{align}
where $b_1$ will be specified later.
If we choose a code with a number of codewords $\tilde M$ that satisfies
\begin{align}
\log \tilde M &\triangleq  C \left( l' - g(C l') \right) - \log l'
\end{align}
we have $(\tilde M-1)\exp\left\{-\gamma\right\} \leq {1}/{l'}$.
Furthermore, by Remark~\ref{rem:prob_error}, the average probability of error is upper-bounded by
\begin{align}
&  \sprob+(1-\sprob)(\tilde M-1)\exp\left\{-\gamma_{k}\right\}  \nonumber\\
&\qquad \leq \frac{l' \epsilon -1}{ l' - 1} + \frac{l'(1-\epsilon) }{l'-1} \frac{1}{l'} = \epsilon.
\end{align}
Suppose it can be shown that
\begin{align}
  \E{\maxx{\tau_1,\tau_2}} \leq l'\label{eq:Emaxtau_U0}
\end{align}
for sufficiently large $l'$. Then the average blocklength is
\begin{align}
(1-\sprob)\E{\maxx{\tau_1,\tau_2} }\leq  \frac{l'(1-\epsilon) }{l'-1}l' \triangleq l.
\end{align}
Consequently, by Theorem~\ref{thm:simple_achiev}, there exists an $(l, M, \epsilon)$-VLSF code with
\begin{align}
  \log M &\geq \log \tilde M \\
  & = C \left( l' - g(C l') \right) - \log l'\\
  &=\frac{C l}{1-\epsilon} - \sqrt{\frac{V_1+V_2}{2\pi(1-\epsilon)}} \sqrt{l}  - \mathcal{O}(l^{\frac{r+1}{4r+2}} \log l)
\end{align}
where the last step follows because
\begin{align}
  l = \frac{(l')^2 (1-\epsilon)}{l'-1} = l' (1-\epsilon) + o(1).
\end{align}

To establish \eqref{eq:Emaxtau_U0}, we proceed as follows. Let $W_n = \imath_{P, W_1}(X_n ; Y_{1,n})$ and $Z_n = \imath_{P,W_2}(X_n ; Y_{2,n})$. We can then upper-bound $\E{\maxx{\tau_1,\tau_2}}$ using the following lemma, \ifthenelse{\boolean{arxiv}}{which is proved in Appendix~\ref{app:two_dim}.}{which is proved using asymptotic results for random walks \cite{Gut2009} and a Berry-Esseen-type theorem that holds for random summations (see proof in \cite[App.~III]{Trillingsgaard2015arxiv}).}
\begin{lemma}
\label{lem:two_dim}
Let $\{W_n\}$ and $\{Z_n\}$, $n\geq 1$, be i.i.d. discrete RVs with $(W_1,Z_1)\sim P_{W,Z}$, positive mean $\mu_W\triangleq\E{W_1}$ and $\mu_Z\triangleq\E{Z_1}$, respectively, and finite moments of order $r\geq 3$, i.e., $\E{|W_1|^r}< \infty$, and $\E{|Z_1|^r}< \infty$. Define the random walks $U_n \triangleq \sum_{i=1}^n W_i$ and $V_n \triangleq \sum_{i=1}^n Z_i$, and the stopping times $\tau_1 \triangleq \inff{n \geq 0: U_n \geq \gamma}$ and $\tau_2 \triangleq \inff{n \geq 0: V_n \geq \gamma}$ for every $\gamma \in\mathbb{R}$.
Then
  \begin{align}
   \E{\maxx{\tau_1, \tau_2}} &\leq \frac{\gamma}{\minn{\mu_W,\mu_Z}}+ \frac{\sigma }{\sqrt{2\pi}}\sqrt{\frac{\gamma}{\mu_W}}\indi{\mu_W=\mu_Z}\nonumber\\
   &\quad+\mathcal{O}\farg{\gamma^{\frac{r+1}{4r+2}}\log \gamma}\label{eq:Etau_upper1}
  \end{align}
  as  $\gamma\rightarrow \infty$,   where $\sigma^2 \triangleq \Va{\frac{W_1}{\mu_W}-\frac{Z_1}{\mu_Z}}$.
\end{lemma}
\spacing

Lemma~\ref{lem:two_dim} implies that there exists a constant $b_1$ such that
\begin{align}
 \E{\maxx{\tau_1(\gamma), \tau_2(\gamma) }} &\leq \frac{\gamma}{C}+g(\gamma)\label{eq:asymp_proof_Emax2}
\end{align}
for sufficiently large $\gamma$.
The conditional average blocklength of the VLSF code can be bounded as follows
\begin{align}
  &\E{\maxx{\tau_1,\tau_2}} = \E{\maxx{\tau_1(\gamma),\tau_2(\gamma)}}\\
  &\leq \frac{\gamma}{C}+g(\gamma)\label{eq:asymp_Emax_final1}  \\
    &=l' -  g(C l') + g(C l'- C g(C l')) \leq l'.\label{eq:asymp_Emax_final2}
\end{align}
Here, \eqref{eq:asymp_Emax_final1} holds by \eqref{eq:asymp_proof_Emax2}, and \eqref{eq:asymp_Emax_final2} follows by the definition of $\gamma$ in \eqref{eq:gamma_def} and the fact that $g(x)$ is nonnegative and nondecreasing.

\bibliographystyle{IEEEtran}
\bibliography{library}

\ifthenelse{\boolean{arxiv}}{
  
\newpage
\appendices

\section{Steps Omitted in the Proof of the Converse Bound}
\label{app:converse_bound}
\subsection{Proof of Property~\ref{prop:VLSF_t}}
\label{app:conve_property}
Let $(f_n,g_{1,n},g_{2,n}, \tau_1, \tau_2, U)$ be a tuple defining an $(l, M, \epsilon)$-VLSF code with $\pr{\maxx{\tau_1,\tau_2} \leq t}\leq \alpha$. Set
  \begin{align}
  \tilde \tau_k &= \left\{ \begin{array}{ll}
t, & \tau_k \leq t  \\
\tau_k, & \tau_k > t
  \end{array}
    \right.
    \end{align}
    and
    \begin{align}
    \tilde g_{k,n}(u, y_k^n) &=  \left\{ \begin{array}{ll}
g_{k,n}(u,y_k^{\tau_k} ), & \tau_k \leq n  \\
g_{k,n}(u,y_k^n), & \tau_k > n.
  \end{array}
    \right..
  \end{align}
  Note that $\tilde \tau_k$ is also a stopping time with respect to the filtration $\mathcal{F}\farg{U, Y_{k}^n}$ for $k\in\{1,2\}$. Since $\tau_k$ is a function of $U$ and $Y_k^{n}$ given $\tau_k\leq n$, the new decoder $\tilde g_{k,n}$ is well-defined. Moreover, the decoders $g_{k,n}$ and $\tilde g_{k,n}$ yield the same probability of error. Thus $(f_n, \tilde g_{1,n}, \tilde g_{2,n}, \tilde \tau_1,\tilde \tau_2,U)$ defines an $(l',M,\epsilon)$-VLSF code, with $l'\geq l$.

\subsection{Proof of \eqref{eq:conve_final_const}}
\label{app:conve_meta_converse}
For each decoder $k$, the average probability of error is no larger than $\epsilon^{(u)}_k$ under $\mathbbm{P}^{(k,u)}_{\bar Y_k, \X}$ and it is no larger than $1-1/M$ under $\mathbbm{Q}^{(k,u)}_{\bar Y, \X}$.
Hence, using the meta-converse theorem \cite[Th.~27]{Polyanskiy2010b} and the inequality \cite[Eq. (102)]{Polyanskiy2010b}, we conclude that
\begin{align}
  \log M &\leq \log \tilde \gamma^{(u)}_k \nonumber\\
  &\quad- \log\Big(\mathbbm{P}^{(k,u)}_{\bar Y_k, \X}\mathopen{}\left[ \imath_k^{(u)}( \X ; \bar Y_k )  \leq \log \tilde \gamma^{(u)}_k\right]-\epsilon_k^{(u)} \Big) \label{eq:converse_logM2}
\end{align}
for all $\tilde \gamma^{(u)}_k$ such that $\mathbbm{P}^{(k,u)}_{\bar Y_k, \X}\mathopen{}\left[ \imath_k( \X ; \bar Y_k )  \leq \log \tilde\gamma^{(u)}_k\right] > \epsilon_k^{(u)}$. Here,
\begin{align}
 \imath_k^{(u)}( \x;\bar  y_k ) &\triangleq \log \frac{\mathbbm{P}^{(k,u)}_{\bar Y_k, \X}(\bar y, \x)}{\mathbbm{Q}^{(k,u)}_{\bar Y_k, \X}(\bar y, \x)} =\log \frac{\mathbbm{P}^{(k,u)}_{\bar Y | \X}(\bar y_k|\x)}{\mathbbm{Q}^{(k,u)}_{\bar Y}(\bar y_k)}
\end{align}
for all $\x\in\mathcal{X}^\infty$ and all $\bar y_k \in\mathcal{Y}_k^{(u)}$. Let now $\varepsilon_{k,M}^{(u)} = \epsilon_k^{(u)} + (\log M)^{-1}$ and set $\tilde \gamma^{(u)} = \gamma_k^{(u)}$ where
\begin{align}
  \gamma_k^{(u)} \triangleq \supp{\nu\in\mathbb{R}: \mathbbm{P}_{\bar Y,\X}^{(k,u)}\mathopen{}\left[\imath_k^{(u)}(\X; \bar Y_k) \leq \log \nu \right] \leq \varepsilon_{k,M}^{(u)}}.\label{eq:converse_gamma2}
\end{align}

Note that there exists an arbitrary small positive constant $\delta$, which is independent of $\log M$, such that
\begin{align}
  & \mathbbm{P}_{\bar Y, \X}^{(k,u)}\mathopen{}\left[\imath_k^{(u)}(\X;\bar Y_k)\leq \log \gamma_k^{(u)} -\delta\right] \nonumber\\
  &\qquad\leq \varepsilon_{k,M}^{(u)} \leq  \mathbbm{P}_{\bar Y ,\X}^{(k,u)}\mathopen{}\left[\imath_k^{(u)}(\X; \bar Y_k) \leq \log \gamma_k^{(u)}\right].\label{eq:converse_delta_properties2}
\end{align}
Using \eqref{eq:converse_gamma2} in \eqref{eq:converse_logM2}, we obtain
\begin{align}
 \log M   &\leq  \log \gamma^{(u)}_k \nonumber\\
 &\quad- \log\left(\mathbbm{P}_{\bar Y,\X}^{(k,u)}\mathopen{}\left[\imath_k^{(u)}(\X; \bar Y_k) \leq \log \gamma^{(u)}_k\right]-\epsilon_k^{(u)} \right) \\
 &\leq  \log \gamma_k^{(u)} + \log\log M.\label{eq:conve_logM_loglogM}
\end{align}
Finally, by \eqref{eq:converse_delta_properties2} and \eqref{eq:conve_logM_loglogM}, we have
\begin{align}
  & \mathbbm{P}_{\bar Y,\X}^{(k,u)}\mathopen{}\left[\imath_k^{(u)}(\X ; \bar Y_k) \leq \log M - \log\log M -\delta \right]\nonumber\\
   &\qquad\leq \mathbbm{P}_{\bar Y,\X}^{(k,u)}\mathopen{}\left[\imath_k^{(u)}(\X ; \bar Y_k) \leq \log  \gamma_k^{(u)} -\delta \right] \\
  &\qquad\leq  \varepsilon_{k,M}^{(u)}.\label{eq:conve_const_before_tildeinf}
\end{align}

Using \cite[Lem.~3]{Tomamichel2013} and the fact that $\mathbbm{Q}^{(k,u)}_{\bar Y}$ is a convex combination of distributions, we obtain the following relation between $\imath_{k}^{(u)}(\x;\bar y)$ and $\tilde \imath_{k}^{(u)}(\x;\bar y)$
\begin{align}
  \imath_{k}^{(u)}(\x;\bar y) \leq \tilde \imath_{k}^{(u)}(\x;\bar y)- \log \frac{1}{|P_t(\mathcal{X})|}.\label{eq:convex_imath_rel}
\end{align}

The inequality in \eqref{eq:conve_const_before_tildeinf} can then be rewritten using \eqref{eq:convex_imath_rel}, as follows:
\begin{align}
 \varepsilon_{k,M}^{(u)}  &\geq \mathbbm{P}_{\bar Y,\X}^{(k,u)}\mathopen{}\left[\tilde \imath_{k}^{(u)}(\X ;\bar Y_k) \leq \log M - \log\log M -\delta\right. \nonumber\\
 &\qquad\qquad\qquad\qquad\qquad\qquad\qquad\left. - \log |\mathcal{P}_{t}\farg{\mathcal{X}}|\right]\\
 &\geq  \mathbbm{P}_{\bar Y,\X}^{(k,u)}\mathopen{}\left[\tilde \imath_{k}^{(u)}(\X ; \bar Y_k) \leq \lambda_t\right].\label{eq:conve_final_const_app}
\end{align}
Here, \eqref{eq:conve_final_const_app} follows by the definition of $\lambda_t$ in \eqref{eq:conve_lambda_def}, and because the number of types $|\mathcal{P}_{t}\farg{\mathcal{X}}|$ is upper bounded by $(t+1)^{|\mathcal{X}-1|}$ \cite[Lem. 1.1]{Tan2014}.

\section{Steps Omitted in the Asymptotic Analysis of the Converse Bound}
\label{app:asymp_converse_bound}

We will need the following property, whose proof follows from standard algebraic manipulations.
\begin{property}
Fix arbitrary $x\in\mathbb{R}$, $a>0, b>0$, and $\lambda >0$. Suppose that $\xi>0$ is the unique solution to the equation
\begin{align}
\frac{\lambda - \xi a}{\sqrt{b \xi}} = x.
\end{align}
Then
\begin{align}
  0 \leq  \xi - \left(\frac{\lambda}{a}  - x \sqrt{\frac{b \lambda}{a^3}} \right)\leq \frac{b}{a^2} x^2.\label{eq:sqrt_property}
\end{align}
\end{property}
\spacing

For notational convenience, we will denote the mean, variance and third absolute moment of $\imath_k(\x^t;Y^t_k)$ by
\begin{align}
  I_{k}(P_{\x^t}) &\triangleq  I(P_{\x^t},W_k) \\
  V_{k}(P_{\x^t}) &\triangleq  V(P_{\x^t},W_k) \\
  T_{k}(P_{\x^t}) &\triangleq  T(P_{\x^t},W_k).
\end{align}

According to \eqref{eq:sqrt_property} and since $\beta$ satisfies \eqref{eq:beta_def}, we have
\begin{align}
  0 \leq \beta - \left( \frac{\lambda}{C} + \tilde Q^{-1}(\epsilon)\sqrt{\frac{V \lambda}{C^3}} \right) \leq \const.
\end{align}

\subsection{Proof of \eqref{eq:conve_sum_hoeffding_bound} and \eqref{eq:conve_sum_hoeffding_bound_2}}
\label{app:asymp_conve_hoeffding}
For the case $t< [0,\alpha)$, we use the following large-deviation bound
\begin{align}
  &\max_{\x^t \in\mathcal{X}^t}\pr{\imath_k(\x^t ; Y_k^t) > \lambda}\nonumber\\
  &\leq \max_{\x^t \in\mathcal{X}^t}\pr{\frac{\imath_k(\x^t;Y^t_k)}{t}-  I_{k}(P_{\x^t})\geq \frac{\lambda}{t} - I_{k}(P_{\x^t})} \\
  &\leq \max_{\x^t \in\mathcal{X}^t} \ee{-\const \left( \frac{\lambda -t I_{k}(P_{\x^t})}{\sqrt{t}} \right)^2}\label{eq:conve_hoeffding}\\
  &\leq  \ee{ -\const   \log^2 \lambda }\label{eq:conve_hoeffding_max_over_x}\\
  &\leq  \left(\frac{1}{\lambda}\right)^{\const  \log \lambda} \label{eq:conve_hoeffding_bound_for_small_t}
\end{align}
where \eqref{eq:conve_hoeffding} follows from Hoeffding's inequality \cite[Th.~2]{Hoeffding1963} and \eqref{eq:conve_hoeffding_max_over_x} follows because $t<\alpha$ and because $I_{k}(P_{\x^t})$ is uniformly upper bounded by $C$.
It follows from \eqref{eq:conve_hoeffding_bound_for_small_t} that
\begin{align}
  &\sum_{t=0}^{\lfloor \alpha \rfloor} \max_{\x^t \in\mathcal{X}^\infty}\pr{\imath_k(\x^t ; Y_k^t) > \lambda}\nonumber\\
  &\qquad \leq (\alpha+1)   \left(\frac{1}{\lambda}\right)^{\const \log \lambda} \\
  &\qquad \leq \const\left(\frac{1}{\lambda}\right)^{\const \log \lambda-1} = o(1).
\end{align}
Using similar argument, one establishes \eqref{eq:conve_sum_hoeffding_bound_2}.

\subsection{Proof of \eqref{eq:conve_sum_berry_esseen_Q} and \eqref{eq:conve_sum_berry_esseen_Q2}} 
\label{app:asymp_conve_berry}
For the case when $t\in [\alpha, \beta)$, we need tighter bounds on $I_{k}(P_{\x^t})$ and $V_{k}(P_{\x^t})$. Let $\Pi_\mu$ be the set of probability distributions that are at distance no larger than $\mu$ from $P^*$:
\begin{align}
  \Pi_{\mu} \triangleq \big\{ P \in \mathcal{P}(\mathcal{X}): \vectornorm{P - P^*}_2\leq \mu \big\}.
\end{align}
Here, $\vectornorm{P - P^*}_2^2 \triangleq \sum_{x \in \mathcal{X}} (P(x) - P^*(x))^2$.
Bounds on $I_{k}(P_{\x^t})$ and $V_{k}(P_{\x^t})$ are then supplied by \cite[Lem.~7]{Tomamichel2013}, which yields positive constants $\varsigma$, $\mu$ and $\rho$ for which
\begin{align}
  I_{k}\farg{P_{\x^t}} &\leq C - \varsigma \vectornorm{P_{\x^t} - P^*}_2^2\label{eq:Dt_bound}\\
    V_k(P_{\x^t}) &\geq \frac{ V_k}{2}\label{eq:Vk_lower_bound}
\end{align}
and
\begin{align}
  \big|\sqrt{V_{k}(P_{\x^t})}-\sqrt{V_k}\big|&\leq \rho \vectornorm{P_{\x^t} - P^*}_2\label{eq:Vkt_bound}
\end{align}
for all $P_{\x^t} \in \Pi_{\mu}$.

Let $P_{\x^t}\in\Pi_{\mu}$. The Berry-Esseen central limit theorem yields the following estimate
\begin{align}
&\pr{\imath_k(\x^t ; Y_k^t) > \lambda}\nonumber\\
&\qquad\leq Q\farg{\frac{\lambda-t I_{k}(P_{\x^t})}{\sqrt{t V_{k}(P_{\x^t})}}}+ \frac{6 t T_{k}(P_{\x^t}) }{(t V_{k}(P_{\x^t}))^{3/2}}\\
&\qquad\leq Q\farg{\frac{\lambda-t I_{k}(P_{\x^t})}{\sqrt{t V_{k}(P_{\x^t})}}}+ \frac{\const}{\sqrt{t}}\label{eq:conve_berry_app}
\end{align}
where the last inequality follows from \eqref{eq:Vk_lower_bound} and because $T_k(P_{\x^t})<\const$ uniformly in $\Pi_\mu$.

This also implies that for all $P_{\x^t} \in\Pi_\mu$,
\begin{align}
& \max_{\x^t \in\mathcal{X}^t}\left\{ \pr{\imath_1(\x^t ; Y_1^t) > \lambda}\right\} \max_{\x^t \in\mathcal{X}^t} \left\{\pr{\imath_2(\x^t ; Y_2^t) > \lambda} \right\}\nonumber\\
&\qquad\qquad\quad\leq \prod_{k=1}^2 \max_{\x^t \in\mathcal{X}^t} Q\farg{\frac{\lambda-t I_{k}(P_{\x^t})}{\sqrt{t V_{k}(P_{\x^t})}}} + \frac{\const}{\sqrt{t}}.\label{eq:conve_berry_esseen_estimate_Q2}
\end{align}

For the case when $P_{\x^t}\not\in\Pi_\mu$, we use Chebyshev's inequality to obtain the estimate
\begin{align}
  \pr{\imath_k(\x^t ; Y_k^t) > \lambda} &\leq
    \frac{t V_{k}\farg{P_{\x^t}}}{(\lambda - t I_{k}\farg{P_{\x^t}})^2}\label{eq:conve_chebyshev}
\end{align}
for all $\lambda > t I_{k}\farg{P_{\x^t}}$. Since $P_{\x^t} \not \in \Pi_{\mu}$,  there exists a constant $C'$ such that $I_{k}\farg{P_{\x^t}} \leq C' < C$. Hence, for sufficiently large $\lambda$, the condition $t\leq \beta$ implies that $\lambda> t I_{k}(P_{\x^t})$. Therefore, by \eqref{eq:conve_chebyshev}, we have that
\begin{align}
&\max_{P_{\x^t} \not \in \Pi_{\mu}}\pr{\imath_k(\x^t ; Y_k^t) > \lambda} \nonumber\\
&\qquad\qquad\leq\max_{P_{\x^t} \not \in \Pi_\mu}\frac{t V_{k}\farg{P_{\x^t}}}{(\lambda - t I_{k}\farg{P_{\x^t}})^2}\\
&\qquad\qquad\leq \frac{ \const t }{(\lambda - t C')^2}\\
&\qquad\qquad\leq \frac{ \const \lambda  }{(\lambda - \lambda C'/C - \const\sqrt{\lambda} - \const)^2}\label{eq:conve_chebyshev2}\\
&\qquad\qquad\leq \frac{\const}{\lambda}\label{eq:chebyshev_bound_final}
\end{align}
where we have used that $t\leq \beta \leq 2t$ for sufficiently large $\lambda$ and that $V_k\farg{P_{\x^t}}$ is uniformly upper-bounded \cite[pp. 7048]{Tomamichel2013}.
We see that $\max_{P_{\x^t} \not \in \Pi_\mu}\pr{\imath_k(\x^t ; Y_k^t) > \lambda}$ can be driven arbitrarily close to zero by having $\lambda$ sufficiently large. This implies that we only need to consider the input vectors $\x^t$ for which $P_{\x^t} \in\Pi_{\mu}$, i.e.,
\begin{align}
  &\max_{\x^t \in\mathcal{X}^t}\pr{\imath_k(\x^t ; Y_k^t) > \lambda} \nonumber\\
  &\qquad\leq \max_{P_{\x^t}  \in \Pi_{\mu}}\pr{\imath_k(\x^t ; Y_k^t) > \lambda} + \frac{\const}{\lambda}.\label{eq:conve_not_in_Pi_mu}
\end{align}
Using \eqref{eq:conve_berry_app} and \eqref{eq:conve_not_in_Pi_mu}, we obtain
\begin{align}
  &\max_{\x^t \in \mathcal{X}^t} \pr{\imath_k(\x^t ; Y_k^t) > \lambda}\nonumber\\
  &\leq \max_{P_{\x^t} \in\Pi_{\mu}} Q\farg{\frac{\lambda-t I_{k}(P_{\x^t})}{\sqrt{t V_{k}(P_{\x^t})}}} + \frac{\const}{\sqrt{t}} +\frac{\const}{\lambda} \\
    &\leq  Q\farg{\min_{P_{\x^t} \in\Pi_{\mu}}\frac{\lambda-t I_{k}(P_{\x^t})}{\sqrt{t V_{k}(P_{\x^t})}}} + \frac{\const}{\sqrt{t}}+\frac{\const}{t} \\
  &= \int_{-\infty}^\infty \phi(x)\indi{\min_{P_{\x^t} \in\Pi_{\mu}}\frac{\lambda-t I_{k}(P_{\x^t})}{\sqrt{t V_{k}(P_{\x^t})}}\leq z} \mathrm{d}z+  \frac{\const}{\sqrt{t}}\label{eq:conve_indi}
\end{align}
for all sufficiently large $\lambda$.
The indicator function in \eqref{eq:conve_indi} can be upper bounded as
\begin{align}
  &\indi{ \min_{P_{\x^t} \in\Pi_{\mu}}\left\{\frac{\lambda-t I_{k}(P_{\x^t})}{\sqrt{t V_{k}(P_{\x^t})}} - z\right\}\leq 0} \nonumber\\
& = \indi{ \max_{P_{\x^t} \in\Pi_{\mu}}\left\{t I_{k}(P_{\x^t})+z\sqrt{t V_{k}(P_{\x^t})}-\lambda \right\}\geq 0}\label{eq:conve_indi1}\\
& \leq \indi{ t C - t \varsigma\xi^2+z\sqrt{t V_k} + |z| \sqrt{t}\rho \xi-\lambda \geq 0}\label{eq:conve_indi2}\\
& \leq \indi{ t C +z\sqrt{t V_k} + \frac{|z| \rho}{2\varsigma}-\lambda \geq 0}\label{eq:conve_indi3}\\
& \leq \indi{ \frac{\lambda -\frac{|z| \rho}{2\varsigma} - t C }{\sqrt{t V_k}} \leq z } \label{eq:conve_indi4}\\
& \leq \indi{ \frac{\lambda}{C} - z\sqrt{\frac{\lambda V_k}{C^3}} - \frac{|z| \rho}{2 C \varsigma} \leq t}\label{eq:conve_indi_final}
\end{align}
where \eqref{eq:conve_indi1} follows since $\sqrt{t V_{k}\farg{\x^t}}>0$ for $P_{\x^t} \in\Pi_{\mu}$ by \eqref{eq:Vk_lower_bound}, \eqref{eq:conve_indi2} follows by \eqref{eq:Dt_bound} and \eqref{eq:Vkt_bound} with $\xi\triangleq\vectornorm{P_{\x^t} - P^*}_2$, \eqref{eq:conve_indi3} follows because $- \varsigma \xi^2 t+ |z|\rho \xi \sqrt{t}$ is a quadratic expression in $\xi\sqrt{t}$ with maximum $\frac{|z|\rho}{2\varsigma}$ and \eqref{eq:conve_indi_final} follows from \eqref{eq:sqrt_property}. The steps \eqref{eq:conve_indi1}--\eqref{eq:conve_indi3} essentially follow from \cite[Prop.~8]{Tomamichel2013}. Substituting \eqref{eq:conve_indi_final} into \eqref{eq:conve_indi} and summing from $(\lfloor \alpha \rfloor+1)$ to $\lfloor \beta \rfloor$, we obtain
\begin{align}
    &\sum_{t=\lfloor \alpha \rfloor+1}^{\lfloor \beta \rfloor}\max_{\x^t \in \mathcal{X}^t } \pr{\imath_k(\x^t ; Y_k^t) > \lambda}\nonumber\\
    &\leq\sum_{t=0}^{\lfloor \beta \rfloor} \int_{-\infty}^{\infty} \phi(z) \indi{ \frac{\lambda}{C} - z\sqrt{\frac{V_k \lambda }{C^3}} - \frac{|z| \rho}{2C \varsigma}\leq t}\mathrm{d}z\nonumber\\
    &\quad+ \mathcal{O}(\log \lambda)\label{eq:conve_central_sum_pr1}\\
     &\leq\int_{0}^{ \beta} \int_{-\infty}^{\infty} \phi(z) \indi{ \frac{\lambda}{C} - z\sqrt{\frac{V_k \lambda}{C^3}} - \frac{|z| \rho}{2C \varsigma} \leq t}\mathrm{d}z\ \mathrm{d}t\nonumber\\
    &\quad + \mathcal{O}(\log \lambda)\label{eq:conve_central_sum_pr2}\\
  &\leq \int_{-\infty}^\infty \phi(z) \int_{0}^{ \beta} \indi{ \frac{\lambda}{C} - z\sqrt{\frac{V_k \lambda}{C^3}} -\frac{|z| \rho}{2C \varsigma} \leq t}\mathrm{d}t \ \mathrm{d}z\nonumber\\
    &\quad + \mathcal{O}(\log \lambda)\label{eq:tonelli}\\
    &\leq \beta-\E{\minn{  \beta ,  \left(\frac{\lambda}{C} - Z_k\sqrt{\frac{V_k \lambda}{C^3}} \right)}} + \mathcal{O}(\log \lambda) \label{eq:sqrt_property_used}\\
        &\leq \sqrt{\frac{V\lambda}{C^3}}\left(\tilde Q^{-1}\farg{\epsilon} - \E{\minn{ \tilde Q^{-1}\farg{\epsilon} , \varrho_k Z_k }}\right) + \mathcal{O}(\log \lambda)\label{eq:conve_sum_berry_esseen_Q_app}
\end{align}
where $\varrho_k$ are defined in Theorem~\ref{thm:asymp} and $Z_k\sim \mathcal{N}(0,1)$.
Here, \eqref{eq:conve_central_sum_pr2} follows because the indicator function is nondecreasing in $t$, in \eqref{eq:tonelli} the order of the integrals is interchangeable  by Tonelli's theorem, and in \eqref{eq:sqrt_property_used} we have used  \eqref{eq:sqrt_property}.

By following the same approach, we obtain \eqref{eq:conve_sum_berry_esseen_Q2}.

\section{Proof of Lemma~\ref{lem:two_dim}}
\label{app:two_dim}
Fix $\gamma\in\mathbb{R}$. We define the following two random walks, which are equivalent to $U_n$ and $V_n$, but more convenient to analyze:
\begin{align}
  A_n &\triangleq U_n/\mu_W + V_n/\mu_Z\label{eq:An}\\
   B_n &\triangleq  U_n/\mu_W -  V_n/\mu_Z.\label{eq:Bn}
\end{align}
We also define the additional stopping time
\begin{align}
 \tau_{12}&\triangleq\inff{n\geq 0 : A_n \geq \gamma \frac{\mu_W+\mu_Z}{\mu_W \mu_Z} }.\label{eq:tau12_def}
\end{align}
We shall next show that
\begin{align}
  \E{\maxx{\tau_1, \tau_2}} &\leq \E{\tau_{12} + \tau_1'(\gamma- U_{\tau_{12}})+\tau_2'( \gamma- V_{\tau_{12}})}\label{eq:tau_1}
\end{align}
where $\tau'_1(\cdot)$ and $\tau'_2(\cdot)$ are defined as
\begin{align}
  \tau'_1( \tilde\gamma) &= \inff{n\geq 0: \sum_{i=1}^{n}\tilde  W_{i}\geq \tilde \gamma}\label{eq:tau_prime1}\\
    \tau'_2(\tilde \gamma) &= \inff{n\geq 0: \sum_{i=1}^{n} \tilde Z_{i}\geq \tilde \gamma}\label{eq:tau_prime2}
\end{align}
and where $\{\tilde W_k, \tilde Z_k\}$ are i.i.d. and $(\tilde W_1,\tilde Z_1)\sim P_{W,Z}$ but independent of $W_j,Z_j$ for all $j\in\mathbb{N}$. Note that $\tau_1'$ and $\tau_2'$ are independent of $ U_{\tau_{12}}$ and $ V_{\tau_{12}}$.

To prove \eqref{eq:tau_1}, we use the following argument. At time $\tau_{12}$, we have that $U_{\tau_{12}}/\mu_W + V_{\tau_{12}}/\mu_Z\geq \gamma \frac{\mu_W + \mu_Z}{\mu_W \mu_Z}$. This implies that either $\tau_1 \leq \tau_{12}$ or $\tau_2 \leq \tau_{12}$ (or both) are satisfied. Consider the case $\tau_1 \leq \tau_{12}$ and $\tau_2>\tau_{12}$. To bound $\E{\maxx{\tau_1,\tau_2}}$, we need to characterize the remaining time until the random walk $V_n$ hits the threshold $\gamma$. This time is given by $\min\{n \geq 0: V_{\tau_{12}+n} \geq \gamma\}$, which has the same distribution as \eqref{eq:tau_prime2} computed at $\gamma-V_{\tau_{12}}$. Note also that $\tau'_k(\tilde \gamma)=0$ for every $\tilde \gamma \leq 0$ since we use the convention $\sum_{i=1}^0 (\cdot)=0$. The inequality in \eqref{eq:tau_1} follows because there exist events for which $\maxx{\tau_1,\tau_2}< \tau_{12}$. The case $\tau_2 \leq \tau_{12}$ and $\tau_1 > \tau_{12}$ can be analyzed similarly.

By \cite[Th. 3.9.4]{Gut2009}
(or by Wald's equality when $W_1$ and $Z_1$ have bounded support \cite[Eq. (106)--(107)]{Polyanskiy2011}), we have
\begin{align}
    \frac{\tilde \gamma}{\mu_W}&\leq \E{\tau_1'(\tilde \gamma)} \leq  \frac{\tilde \gamma}{\mu_W} + \const\label{eq:tau1_exp}\\
  \frac{\tilde \gamma}{\mu_W}&\leq \E{\tau_2'(\tilde \gamma)} \leq  \frac{\tilde \gamma}{\mu_Z} + \const\label{eq:tau2_exp}\\
 \gamma \frac{\mu_W+\mu_Z}{2\mu_W \mu_Z} &\leq \E{\tau_{12}}\leq \gamma \frac{\mu_W+\mu_Z}{2\mu_W \mu_Z}+\const .\label{eq:tau12_exp}
\end{align}

Using \eqref{eq:tau_1}, the linearity of expectation, \eqref{eq:tau1_exp}--\eqref{eq:tau12_exp}, and the fact that
\begin{align}
\E{\tau_1'( \gamma -  U_{\tau_{12}})}
&= \E{\E{\tau_1'( \gamma -  U_{\tau_{12}})| U_{\tau_{12}}}}\nonumber\\
&\leq \frac{1}{\mu_W}\E{( \gamma -  U_{\tau_{12}})^+}+\const
\end{align}
we conclude that
\begin{align}
 & \E{\maxx{\tau_1,\tau_2}} - \gamma \frac{\mu_W+\mu_Z}{2\mu_W \mu_Z}\nonumber\\
  &\leq \frac{1}{\mu_W}\E{\left( \gamma-  U_{\tau_{12}}\right)^+} +\frac{1}{\mu_Z}\E{\left( \gamma-  V_{\tau_{12}}\right)^+}+\const\\
  &=\frac{1}{\mu_W}\E{\left( \gamma- \frac{1}{2}\mu_W(A_{\tau_{12}}+ B_{\tau_{12}})\right)^+}\nonumber\\
  &\quad +\frac{1}{\mu_Z}\E{\left( \gamma- \frac{1}{2}\mu_Z\left(A_{\tau_{12}} - B_{\tau_{12}}\right)\right)^+}+\const\\
    &\leq \E{\left( \frac{\gamma}{\mu_W}- \frac{1}{2}\left( \gamma \frac{\mu_W + \mu_Z}{\mu_W \mu_Z}+ B_{\tau_{12}}\right)\right)^+} \nonumber\\
    &\quad+\E{\left( \frac{\gamma}{\mu_Z}- \frac{1}{2}\left( \gamma \frac{\mu_W + \mu_Z}{\mu_W \mu_Z} -  B_{\tau_{12}}\right)\right)^+}+\const\label{eq:achiev_Emax_def_tau12}\\
      &= \frac{1}{2}\E{\left| \gamma \frac{\mu_Z - \mu_W}{\mu_W \mu_Z}- B_{\tau_{12}}\right|}+\const\label{eq:tau_before_clt}
 \end{align}
 where \eqref{eq:achiev_Emax_def_tau12} follows from the definition of $\tau_{12}$ (see \eqref{eq:tau12_def}) which implies that $A_{\tau_{12}}\geq  \gamma \frac{\mu_W + \mu_Z}{\mu_W \mu_Z}$.

We next show that the RHS of \eqref{eq:tau_before_clt} is upper-bounded by the RHS of \eqref{eq:Etau_upper1} by the following two steps. First, we shall approximate $B_{\tau_{12}}$ by a Gaussian RV using a variation of the Berry-Esseen theorem that holds when the number of terms in the summation is a RV (see Lemma~\ref{lem:clt_random_sums} below). Then, we shall establish \eqref{eq:Etau_upper1} using standard properties of Gaussian RVs.
\begin{lemma}{(\cite[Th. 1]{Landers1976})}
\label{lem:clt_random_sums}
  Let $\{\xi_n, n\geq 1\}$ be i.i.d. RVs with zero mean, positive variance $\sigma^2$, and finite third absolute moment. Let $\{N_n, n\in\mathbb{N}\}$ be a sequence of positive integer-valued RVs and assume that
\begin{align}
  \pr{ \left|\frac{N_n}{n\nu} -1\right| > \zeta_{n}} &= \bigoh{\sqrt{\zeta_n}}\label{eq:clt_Nk_concentation}
\end{align}
for some constant $\nu$ and a sequence $\{\zeta_n\}$ that vanishes as $n\rightarrow \infty$ and that satisfies $\frac{1}{n}\leq \zeta_n$ for all $n$.
Then
\begin{align}
 \sup_{\lambda \in \mathbb{R}} \left| \pr{ \sum_{i=1}^{N_n} \xi_i \leq \sigma\sqrt{n\nu} \lambda} - \Phi(\lambda) \right| &= \bigoh{\sqrt{\zeta_n}}.
\end{align}
 \end{lemma}
\spacing

The RV $B_{\tau_{12}}$ and its variance satisfies \cite[Th. 4.2.4 (ii')]{Gut2009}
\begin{align}
  \Va{B_{\tau_{12}}} = \sigma^2 \gamma \frac{\mu_W + \mu_Z}{2\mu_W \mu_Z} +\mathcal{O}(1)\label{eq:Btau12_var}
\end{align}
as $\gamma \rightarrow \infty$.
For some constant $\nu>0$, let $\gamma_n\triangleq\frac{2\nu \mu_W \mu_Z  n}{\mu_W+\mu_Z}$, $N_n \triangleq \tau_{12}( \gamma_n)$, $\xi_n \triangleq \frac{W_n}{\mu_W} - \frac{Z_n}{\mu_Z}$ and $\zeta_n \triangleq n^{-\frac{r}{2r+1}}$ for $n\in\mathbb{N}$. Note that by \eqref{eq:tau12_exp}, we have
\begin{align}
  \E{N_n} &=  \E{\tau_{12}(\gamma_n)} \\
  &= \gamma_n\frac{\mu_W+\mu_Z}{2\mu_W \mu_Z}   +\mathcal{O}(1)\\
    &=\nu n+\mathcal{O}(1),\qquad n\rightarrow \infty.\label{eq:Nk_exp}
\end{align}
%The variance of the stopping time $N_k$ is given by \cite[Th. 3.9.1]{Gut2009}
%\begin{align}
%  \Va{N_k} &= \underbrace{\frac{1}{8}\Va{\frac{W_k}{\mu_W} + \frac{Z_k}{\mu_Z}}}_{c_1}  \nu k + o(k).\label{eq:Nk_var}
%\end{align}
We next show that condition \eqref{eq:clt_Nk_concentation} in Lemma~\ref{lem:clt_random_sums} is satisfied. Indeed,
\begin{align}
&\pr{ \left|\frac{N_n}{\nu n} -1\right|\geq \zeta_{n}}\nonumber\\
 &= \pr{ \left|\frac{N_n-\nu n}{\sqrt{\nu n}} \right|^r\geq \left(\sqrt{\nu n}\zeta_{n}\right)^r} \\ &\leq  \frac{\E{\left|\frac{N_n-\nu n}{\sqrt{\nu }}\right|^r}}{ (\sqrt{\nu n} \zeta_n)^r}\label{eq:Nn_concentration1}\\
 &= \frac{\const}{(\sqrt{\nu n} \zeta_n)^r} \label{eq:Nn_concentration2}\\
&= \frac{\const}{n^{r/2} \left(n^{-\frac{r}{2r+1}}\right)^r} = \frac{\const}{n^{\frac{r}{4r+2}}} = \mathcal{O}(\sqrt{\zeta_n})
\end{align}
as $n\rightarrow \infty$. Here, \eqref{eq:Nn_concentration1} follows from Markov's inequality and \eqref{eq:Nn_concentration2} follows from \cite[Th. 3.8.4(i)]{Gut2009}.

% and (c) follows from \eqref{eq:Nk_exp} and \eqref{eq:Nk_var}.
Let $F(\lambda) \triangleq \pr{B_{N_n}\leq \sigma \sqrt{v n}\lambda}$.
We can now use Lemma~\ref{lem:clt_random_sums}, which for sufficiently large $n$  implies that
\begin{align}
\sup_{\lambda \in \mathbb{R}} \left| F(\lambda)- \Phi(\lambda) \right| &\leq \const n^{-\frac{r}{4r+2}}.\label{eq:CLT_consequence}
\end{align}

We next refine our estimate in \eqref{eq:CLT_consequence} using Lemma~\ref{lem:petrov} below.
\begin{lemma}{(\cite[Th. 9]{V.V.Petro})}
  Let $F(x)$ be the cumulative distribution function of a RV that has finite moment of order $p$. Suppose that $0 < \Delta \triangleq \sup_x |F(x) - \Phi(x)|\leq 1/\sqrt{e}$. Then there exists a constant $C_p$, that depends only on $p$, such that
  \begin{align}
  |F(x)-\Phi(x)| \leq \frac{C_p \Delta  \left(\log \frac{1}{\Delta}\right)^{p/2}+\rho_p}{1+|x|^p}\label{eq:sums_pretov_theorem10}
  \end{align}
  for all $x$. Here
  \begin{align}
  \rho_p = \left|\int_{-\infty}^\infty |x|^p \mathrm{d}F(x) - \int_{-\infty}^\infty |x|^p \mathrm{d}\Phi(x) \right|.
  \end{align}
  \label{lem:petrov}
\end{lemma}
\spacing

Using Lemma~\ref{lem:petrov} and \eqref{eq:CLT_consequence}, we have that
\begin{align}
  \left|F(\lambda) - \Phi(\lambda) \right|   &\leq \frac{\const n^{-\frac{r}{4r+2}}  \log n+\rho_2(n)}{1+\lambda^2}.\label{eq:achiev_nonuniform_bound}
\end{align}
for $\lambda \in\mathbb{R}$ and sufficiently large $n$. Here,
\begin{align}
\rho_2(n)
&= \left|\frac{\Va{B_{N_n}}}{\sigma^2 n \nu} - 1 \right| =  \left|\frac{ n+\mathcal{O}(1) }{ n } - 1 \right| \leq \frac{\const}{n}.\label{eq:rho_asymp}
\end{align}
Fix an arbitrary $a\in\mathbb{R}$.
Using \eqref{eq:achiev_nonuniform_bound}, we obtain the following upper bound
\begin{align}
&\E{\left|a - B_{N_n}\right|} \nonumber\\
&=\sigma \sqrt{\nu n} \int_{0}^\infty 1+F\farg{\frac{a}{\sigma \sqrt{\nu n}} - x} -F\farg{\frac{a}{\sigma \sqrt{\nu n}} + x} \mathrm{d}x\\
&\leq \sigma \sqrt{\nu n} \int_{0}^\infty \left[\Phi\left(\frac{a}{\sigma \sqrt{\nu n}} - x\right) + \left(1-\Phi\left(\frac{a}{\sigma \sqrt{\nu n}} + x\right)\right)\right.\nonumber\\
&\qquad \left. +  \frac{\const n^{-\frac{r}{4r+2}}  \log n+\const/n}{1+(\frac{a}{\sigma\sqrt{\nu n}} + x)^2}+ \frac{\const n^{-\frac{r}{4r+2}} \log n+\const/n}{1+(\frac{a}{\sigma\sqrt{\nu n}} - x)^2}\right]  \mathrm{d}x\\
&= \sigma \sqrt{\nu n} \E{\left|\frac{a}{\sigma \sqrt{\nu n}} - Z\right|}\nonumber\\
&\quad+\pi\sigma \sqrt{\nu}\left(\const n^{\frac{1}{2}-\frac{r}{4r+2}}  \log n+\const/\sqrt{n}\right)\\
&=\sqrt{\frac{2}{\pi}} \sigma \sqrt{\nu n}\psi\left(\frac{a}{\sigma \sqrt{\nu n}}\right) + |a| +\mathcal{O}(n^{\frac{r+1}{4r+2}}\log n)\label{eq:Eabs_bound}
\end{align}
as $n\rightarrow \infty$, where $Z\sim \mathcal{N}(0,1)$ and
\begin{align}
&\psi(x)\triangleq\sqrt{\frac{\pi}{2}}(\E{|x - Z|}-|x|)\\
&=  \exp\left(-\frac{x^2}{2}\right) + x\sqrt{\frac{\pi}{2}} \ \left(\text{erf}\left( \frac{x}{\sqrt{2}} \right) - \text{sgn}(x)\right).
\end{align}
The positive function $\psi(x)$ is unimodal with maximum $1$ attained at $x=0$ and decays exponentially to $0$ as $|x|\rightarrow  \infty$.

Substituting $a = \gamma_n \frac{\mu_Z - \mu_W}{\mu_W\mu_Z}  =\frac{2 \nu n (\mu_Z - \mu_W)}{\mu_W + \mu_Z}$ into \eqref{eq:Eabs_bound}, we obtain
\begin{align}
  &\E{\left|\gamma_n \frac{\mu_Z - \mu_W}{\mu_W\mu_Z} - B_{N_n}\right|} \nonumber\\
  &\leq  \sqrt{\frac{2}{\pi}} \sigma \sqrt{\nu n}\psi\farg{\frac{2 \sqrt{\nu n} (\mu_Z - \mu_W)}{\sigma(\mu_W + \mu_Z)}} \nonumber\\
  &\quad+ \gamma_n\left|\frac{\mu_Z - \mu_W}{\mu_W\mu_Z}\right| +\mathcal{O}(n^{\frac{r+1}{4r+2}}\log n).\label{eq:EBN_k_bound}
\end{align}
Note that for the case $\mu_Z \not= \mu_W$, we have that $ \sqrt{ n}\psi\farg{\frac{2 \sqrt{\nu n} (\mu_Z - \mu_W)}{\sigma(\mu_W + \mu_Z)}}=o(1)$ as $n\rightarrow \infty$. Substituting \eqref{eq:EBN_k_bound} into \eqref{eq:tau_before_clt}, we obtain
\begin{align}
  &\E{\maxx{\tau_1(\gamma_n), \tau_2(\gamma_n)}} \nonumber\\
  &\leq \gamma_n  \frac{\mu_W + \mu_Z}{2\mu_W \mu_Z}\nonumber\\
  &\quad + \frac{1}{2} \E{\left|\gamma_n \frac{\mu_Z - \mu_W}{\mu_W\mu_Z} - B_{\tau_{12}(\gamma_n)}\right|}+\mathcal{O}(1)\\
  &=\gamma_n \frac{\mu_W + \mu_Z}{2\mu_W \mu_Z}+ \gamma_n\left| \frac{\mu_Z - \mu_W}{2\mu_W\mu_Z}\right|\nonumber\\
  &\quad+\frac{\sigma }{\sqrt{2\pi}}\sqrt{\nu n}\indi{\mu_W=\mu_Z}  +\mathcal{O}(n^{\frac{r+1}{4r+2}}\log n)\\
  &=\frac{\gamma_n}{\minn{\mu_W, \mu_Z}}+ \frac{\sigma }{\sqrt{2\pi}}\sqrt{\nu n}\indi{\mu_W=\mu_Z}\nonumber\\
  &\quad+\mathcal{O}(n^{\frac{r+1}{4r+2}}\log n), \qquad n\rightarrow \infty\label{eq:Emax_asymp_intermediate}
\end{align}
where \eqref{eq:Emax_asymp_intermediate} follows from the identity $a+b + |a-b| = 2\maxx{a,b}$.

  To complete the proof, let $n_1 \triangleq \lceil \frac{\gamma}{\min(\mu_W, \mu_Z)} \rceil$, $\Psi\farg{x} \triangleq x + \frac{\sigma}{\sqrt{2\pi}} \sqrt{\nu x}\indi{\mu_W=\mu_Z} + b_1 x^{\frac{r+1}{4r+2}}\log x$, and set $\nu\triangleq\frac{\mu_W + \mu_Z}{2\maxx{\mu_W, \mu_Z}}$, i.e.
  \begin{align}
  \gamma_n  = \minn{\mu_W,\mu_Z}n.
\end{align}
Note that $\Psi(x)$ is nondecreasing, concave and differentiable in $x\in [1,\infty]$.
Then there exists a constant $b_1>0$ such that
 \begin{align}
 &\E{\maxx{\tau_1(\gamma), \tau_2(\gamma)}}\nonumber\\
  &\leq \E{\maxx{\tau_1(\gamma_{n_1}), \tau_2(\gamma_{n_1})}}\label{eq:Emax_final1}\\
 &\leq n_1+ \frac{\sigma }{\sqrt{2\pi}}\sqrt{\nu n_1}\indi{\mu_W=\mu_Z}+b_1 n_1^{\frac{r+1}{4r+2}}\log n_1\\
 &=\Psi\farg{\left\lceil \frac{\gamma}{\min(\mu_W, \mu_Z)} \right\rceil}\\
 &\leq \Psi\farg{\frac{\gamma}{\minn{\mu_W, \mu_Z}}+1}\label{eq:Emax_final3}\\
 &\leq \Psi\farg{\frac{\gamma}{\min(\mu_W, \mu_Z)}} + \const\\
  &= \frac{\gamma}{\minn{\mu_W,\mu_Z}}+ \frac{\sigma }{2\sqrt{\pi}}\sqrt{\frac{\gamma(\mu_W + \mu_Z)}{\mu_W\mu_Z}}\indi{\mu_W=\mu_Z}\nonumber\\
 &\quad{}+\mathcal{O}(\gamma^{\frac{r+1}{4r+2}}\log \gamma)\\
 &= \frac{\gamma}{\minn{\mu_W,\mu_Z}}+ \frac{\sigma }{\sqrt{2\pi}}\sqrt{\frac{\gamma}{\mu_W}}\indi{\mu_W=\mu_Z}\nonumber\\
 &\quad{}+\mathcal{O}(\gamma^{\frac{r+1}{4r+2}}\log \gamma).
 \end{align}
 Here, \eqref{eq:Emax_final1} follows because $\E{\maxx{\tau_1(\gamma), \tau_2(\gamma)}}$ is nondecreasing in $\gamma$.
}{}



%\bibliographystyle{IEEEtran}

\end{document}